# Pulsational pair instability as an explanation for the most luminous supernovae

S. E. Woosley[1], S. Blinnikov[1,2,3] & Alexander Heger[1,4]

[1]Department of Astronomy and Astrophysics, UCSC, Santa Cruz, California 95064, USA. [2]ITEP, 117218 Moscow, Russia. [3]Max Planck Institut für Astrophysik, Garching, D-85741, Germany. [4]Theoretical Astrophysics Group, T-6, MS B227, Los Alamos National Laboratory, Los Alamos, New Mexico 87544, USA.

**The extremely luminous supernova SN 2006gy (ref. 1) challenges the traditional view that the collapse of a stellar core is the only mechanism by which a massive star makes a supernova, because it seems too luminous by more than a factor of ten. Here we report that the brightest supernovae in the modern Universe arise from collisions between shells of matter ejected by massive stars that undergo an interior instability arising from the production of electron–positron pairs[2]. This 'pair instability' leads to explosive burning that is insufficient to unbind the star, but ejects many solar masses of the envelope. After the first explosion, the remaining core contracts and searches for a stable burning state. When the next explosion occurs, several solar masses of material are again ejected, which collide with the earlier ejecta. This collision can radiate $10^{50}$ erg of light, about a factor of ten more than an ordinary supernova. Our model is in good agreement with the observed light curve for SN 2006gy and also shows that some massive stars can produce more than one supernova-like outburst.**

The life of a star is determined by the mass, composition and rotation rate with which it is born. Most important to its death is the mass, at the end of the star's life, of its core of helium and heavy elements, usually called the 'helium core' (Table 1). That mass, which determines the explosion mechanism for the supernova the star makes and its nucleosynthesis, is in turn sensitive to how much mass the star lost along the way. For currently favoured mass loss rates and solar composition, stars with initial masses over ~40 times that of the Sun lose all their hydrogen envelope and part of their helium core as well. Current calculations suggest that the maximum helium core at death is only about 15 solar masses[3,4], and no modern star, at least in our Galaxy, would encounter the pair instability. The rate at which the most massive stars lose mass is quite uncertain[5–7] though, and depends upon their metal content[8,9]. Much more massive helium cores could have been common for stars born in the distant past and perhaps, occasionally, even today.

Among these most massive stars, particularly poorly explored are "pulsational pair-instability supernovae"[2,10,11], which might occur at the deaths of main-sequence stars in the mass range 95 to 130 solar masses (Table 1). The pair instability is encountered when, late in the star's life, a large amount of thermal energy goes into making the masses of an increasing abundance of electron–positron pairs rather than providing pressure. Rapid contraction occurs, followed by a thermonuclear explosion[2,12,13]. But in the pulsational case, the energy released by the explosive burning is inadequate to unbind the entire star. It suffices, however, violently to eject many solar masses of surface material, including all that is left of the hydrogen envelope, in a series of giant 'pulses'. The typical binding energy for the hydrogen envelope of such massive stars is only ~0.1 to $1 \times 10^{49}$ erg, whereas the energy of a pulse is ~$1–100 \times 10^{49}$ erg (Supplementary Table 1), so the envelope is easily ejected in the first pulse. After each pulse, the remaining core contracts, radiates neutrinos and light, and searches again for a stable burning state. The time required for this contraction is sensitive to the strength of the pulse and how close the star came to becoming unbound. If the temperature after the first pulse is less than about $9 \times 10^8$ K, neutrino losses are inefficient and it may be decades before the star starts burning again. If the core is much hotter, it may only take days.

If the remaining helium core is still over 40 solar masses, with the exact threshold depending upon the entropy lost to neutrinos during the interpulse period, the star encounters the instability again, and ejects another several solar masses. Later ejections have lower mass, because the envelope was expelled in the first pulse, but have higher energy. They quickly catch up to the first shell, which by this time is at $10^{15}$–$10^{16}$ cm, where the collision dissipates most of their relative kinetic energy as radiation (Fig. 1). Because of the large radius for the collision, adiabatic losses from expansion are roughly two orders of magnitude less than in a common type-II supernova. That is, a collision involving only $10^{50}$ erg of kinetic energy can radiate as much as $10^{50}$ erg of light, more than ten times an ordinary supernova.

To illustrate these general ideas, consider the evolution of a star of 110 solar masses and solar composition. Its evolution is calculated using the Kepler code[3,14] with mass loss included at a fraction of the standard value for solar metallicity stars[15,16], 50% on the main sequence, 10% as a helium-burning red giant. Our 110-solar-mass main-sequence star then ends its life with a total mass of 74.6 solar masses and a helium core of 49.9 solar masses, well within the pulsational domain. The pre-supernova star is a red supergiant with radius $1.1 \times 10^{14}$ cm and luminosity $9.2 \times 10^{39}$ erg s$^{-1}$. Its outer 24 solar masses of low-density envelope are only bound by $9.0 \times 10^{48}$ erg.

After burning helium and carbon, when the temperature exceeds $10^9$ K, this star first encounters the pair instability. The helium core collapses rapidly to a maximum central temperature of $3.04 \times 10^9$ K and density $1.50 \times 10^6$ gm cm$^{-3}$, far hotter than the usual $2.0 \times 10^9$ K at which oxygen burns stably in a massive star. So the star violently explodes, burning 1.49 solar masses of oxygen and 1.55 solar masses of carbon and releasing $1.4 \times 10^{51}$ erg (Supplementary Figs 1 and 2). Most of this energy goes into expanding the star. About 10%, however, goes into driving off 24.5 solar masses of envelope and core (mostly helium and some hydrogen) with a terminal speed of 100 to 1,000 km s$^{-1}$ (Fig. 1). This envelope ejection gives the first supernova-like display with a luminosity ~$4 \times 10^{41}$ erg s$^{-1}$ for 200 days (Fig. 2 and Supplementary Fig. 5).

What is left behind is a 50.7-solar-mass remnant, slightly larger than the original helium core mass, that once again radiates neutrinos, contracts and grows hotter. Then 6.8 years later, it encounters the pair instability a second time. This time the pulse is stronger, and $6.0 \times 10^{50}$ erg is shared by a smaller ejected mass of 5.1 solar masses. The collision of this high-velocity shell with the larger mass ejected earlier (Fig. 1) produces a brilliant light curve[17] calculated here using the radiation-hydrodynamics code Stella[18] (Fig. 3). Stella uses multi-energy groups to compute the coupling of radiation transfer to the gas dynamics and produces multi-colour and bolometric light curves. Previously Stella was used successfully to resolve a very thin shell and radiative shock in the case of SN 1994W and gave multi-colour fluxes in good agreement with observations[19]. The good agreement with SN 2006gy[1], (Fig. 3 and Supplementary Figs 7–9) is suggestive of a light curve generated by collisions between solar masses of material as refs 1 and 20 have also proposed. Other models for SN 2006gy based upon traditional pair-instability supernovae can be very bright[12,21], but require a large mass of $^{56}$Ni and are difficult to reconcile with the narrow width of the observed light curve[22,23] and the narrow spectral features due to hydrogen. They also require exceptionally massive progenitor stars.

The photospheric structure of these collisionally dominated supernovae is novel. Early on, the collision occurs at such high density and small radius that the shock is preceded by an optically thick photosphere (Supplementary Figs 10 and 11). Matter inside this photosphere is ionized and optically thick. The emission is nearly blackbody and no X-ray or radio emission is produced. The mass in the second eruption quickly becomes concentrated in very thin shells. Later, the large velocity shear bounding these shells keeps the opacity in Doppler-broadened lines from becoming too small, and the emission continues to be predominantly in near-optical bands. The large column depth probably keeps any appreciable X-rays that are produced from escaping until after the optical display is over. This is consistent with the low level of X-rays detected from the supernova[20].

Nine years later, the 110-solar-mass model finishes a final phase of contraction and gently starts silicon burning at its centre, making an iron core that collapses (Fig. 2 and Supplementary Fig. 4). A 95-solar-mass star similarly evolved, but with mild rotation (equatorial speed 100 km s⁻¹ on the main sequence) and magnetic torques[24], produces a similar helium-core mass, but has sufficient angular momentum in its iron core to make a neutron star with period 2 ms. This is sufficiently rapid rotation to form a magnetar[25], or, within uncertainties in the angular momentum transport model, a collapsar[26]. Thus the final death of the star might generate a gamma-ray burst[27,28], but one that is embedded in many solar masses of circumstellar material. The optical light curve from such an event could also be very bright and might be an alternative explanation for SN 2006gy.

Indeed, as Supplementary Table 1 shows, the pulsational–pair instability mechanism can energize a variety of explosive phenomena with characteristic timescales ranging from days to centuries. We have focused here on the brightest of these events, but if the energy of the ejected shells is low and if the core of the star eventually collapses to a slowly rotating black hole, we might observe "supernova impostors"[29] and nothing else. If the star lost its envelope, but retained a supercritical helium core mass, it might form a repeating type Ib supernova[30].

**Supplementary Information** is linked to the online version of the paper at www.nature.com/nature.

**Acknowledgements** This work was supported by the Scientific Discovery through Advanced Computing (SciDAC) Program of the US Department of Energy, by NASA, and by the Russian Foundation for Basic Research and Science Schools. At Los Alamos, this work was carried out under the auspices of the National Nuclear Security Administration of the US Department of Energy.

**Author Contributions** S.E.W. and A.H. proposed that the light curves of pulsational pair instability supernovae might have a large range in luminosity including exceptionally brilliant supernovae. They carried out the calculations of stellar evolution and explosion. S.B. provided expertise in the physics of supernovae with circumstellar interaction and calculated all the light curves from the models except those done with Kepler.

**Author Information** Reprints and permissions information is available at www.nature.com/reprints. Correspondence and requests for materials should be addressed to S.E.W. (woosley@ucolick.org).

**Table 1 Final evolution of stars of different initial mass**[3]

| Mass at birth (solar masses) | Helium core mass (solar masses) | Compact remnant | Event |
|---|---|---|---|
| 10–95 | 2–40 | Neutron star, black hole | Ordinary supernova |
| 95–130 | 40–60 | Neutron star, black hole | Pulsational pair-instability supernova |
| 130–260 | 60–137 | Explosion, no remnant | Pair-instability supernova |
| >260 | >137 | Black hole | ? |

Column 1 gives the total mass of the (non-rotating) star when it is born. If the outer layers of hydrogen and helium are not entirely lost along the way, the second column gives the mass of the core of helium and heavier elements inside the star when it dies. Columns 3 and 4 then describe how the star dies and what sort of remnant it leaves behind. Without rotation, helium cores over 137 solar masses simply disappear into a black hole. With rotation, their evolution is uncertain. The pair instability occurs after carbon burning when the centre of the star encounters thermodynamic conditions where a large fraction of the internal energy is stored in the rest masses of electron–positron pairs. The loss of pressure renders the star briefly unstable against collapse, nuclear burning and explosion.

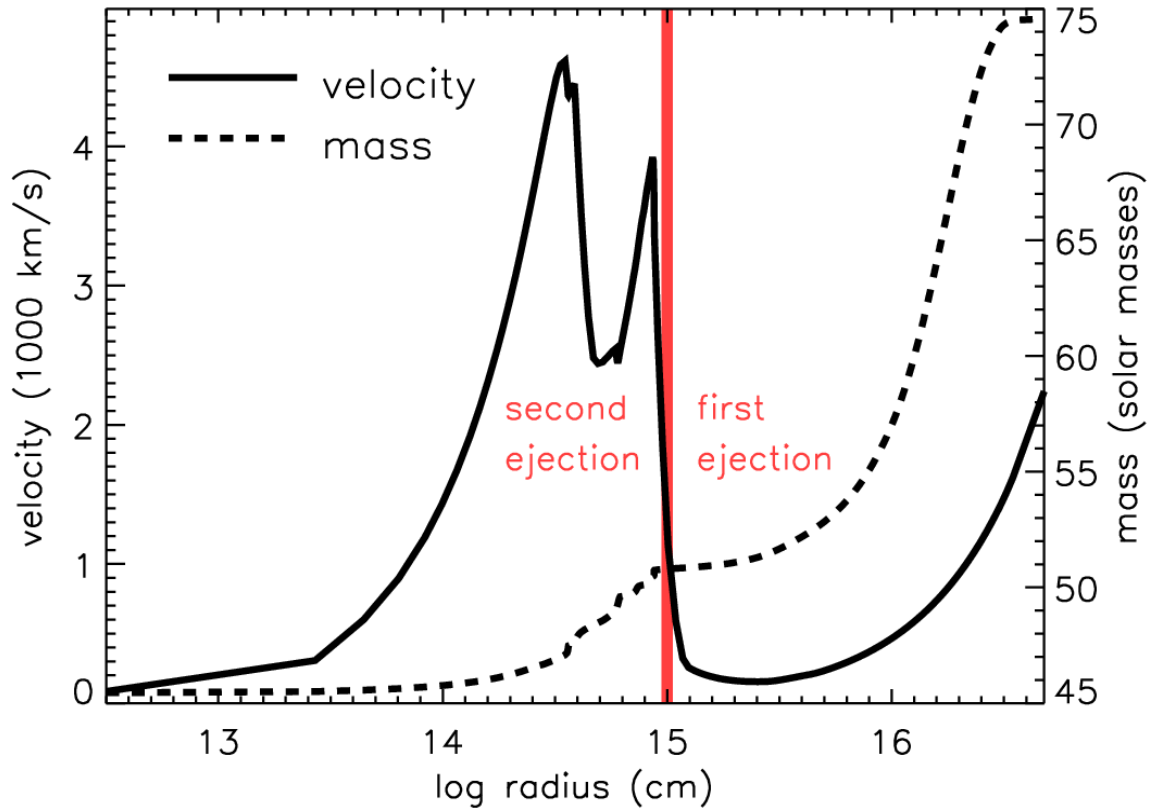

**Figure 1 Velocity structure following the second eruption of a 110-solar-mass pulsational pair-instability supernova.** The velocity and enclosed mass are plotted against the log of the radius. The velocity discontinuity at $10^{15}$ cm shows where fast-moving ejecta from the second outburst are starting to impact the slower-moving material ejected in the first pulse. Hydrogen-rich and helium-rich material immediately above this shock is moving at less than 200 km s[-1] and will give rise to narrow lines in the spectrum of the emission, as was seen in SN 2006gy[1]. Most of the kinetic energy of the second ejection will be dissipated within $10^{16}$ cm. This particular configuration resulted from a star initially with 110 solar masses that had 74.6 solar masses left when it began exploding (see text).

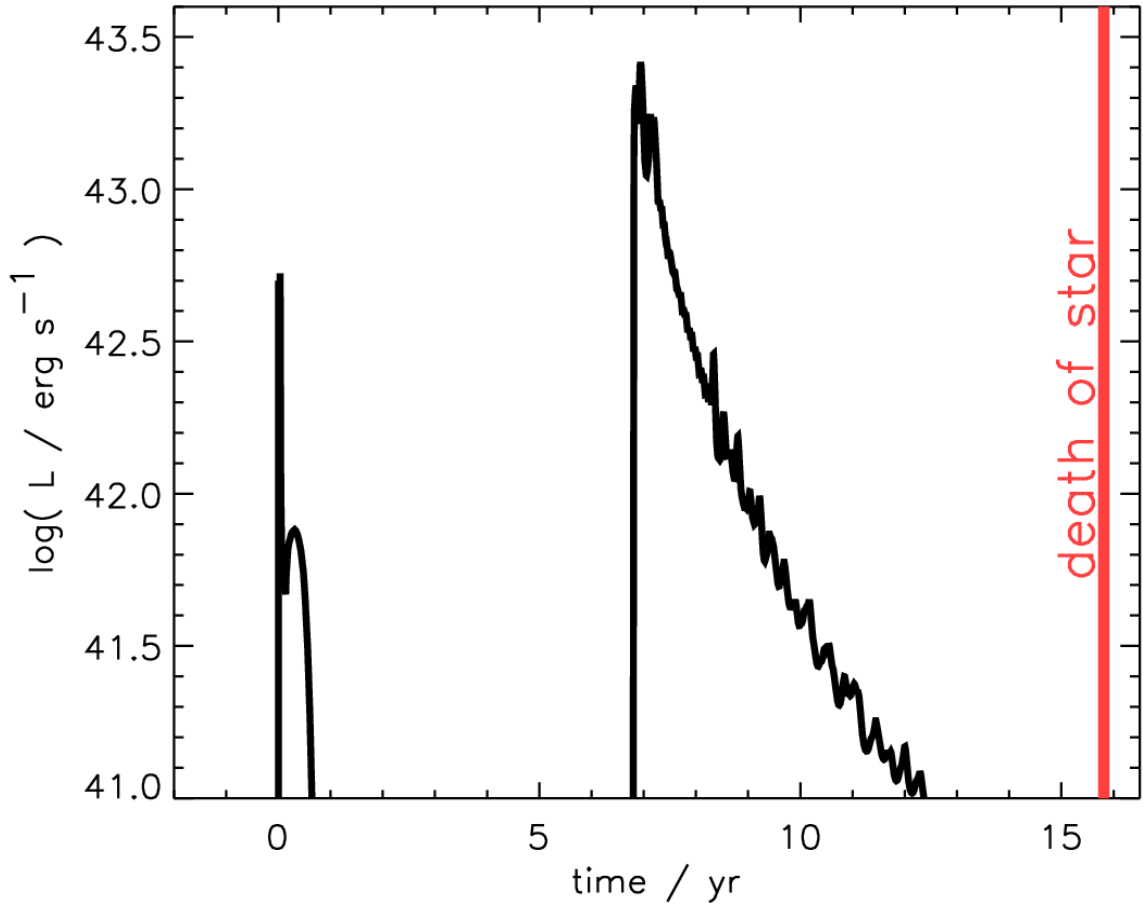

**Figure 2 Cumulative light curve for the 110-solar-mass model.** Three events characterize the final years of the star's life. The first major eruption ejects about 25 solar masses of hydrogen–helium envelope and makes a supernova with luminosity ~6 × 10⁴¹ erg s⁻¹ lasting 200 days (see Supplementary Fig. 5). Shock breakout produces the brief bright ultraviolet transient at the onset of this first light curve, while the plateau is due to hydrogen recombination. Then 6.9 years later a second eruption produces a brilliant event as the fast-moving ejecta collide with the debris of the first supernova (Fig. 1 and 3). And 9 years after that, the star forms a 2.2-solar-mass iron core that collapses to a rapidly rotating neutron star or black hole. A third bright event, possibly a gamma-ray burst, might then occur.

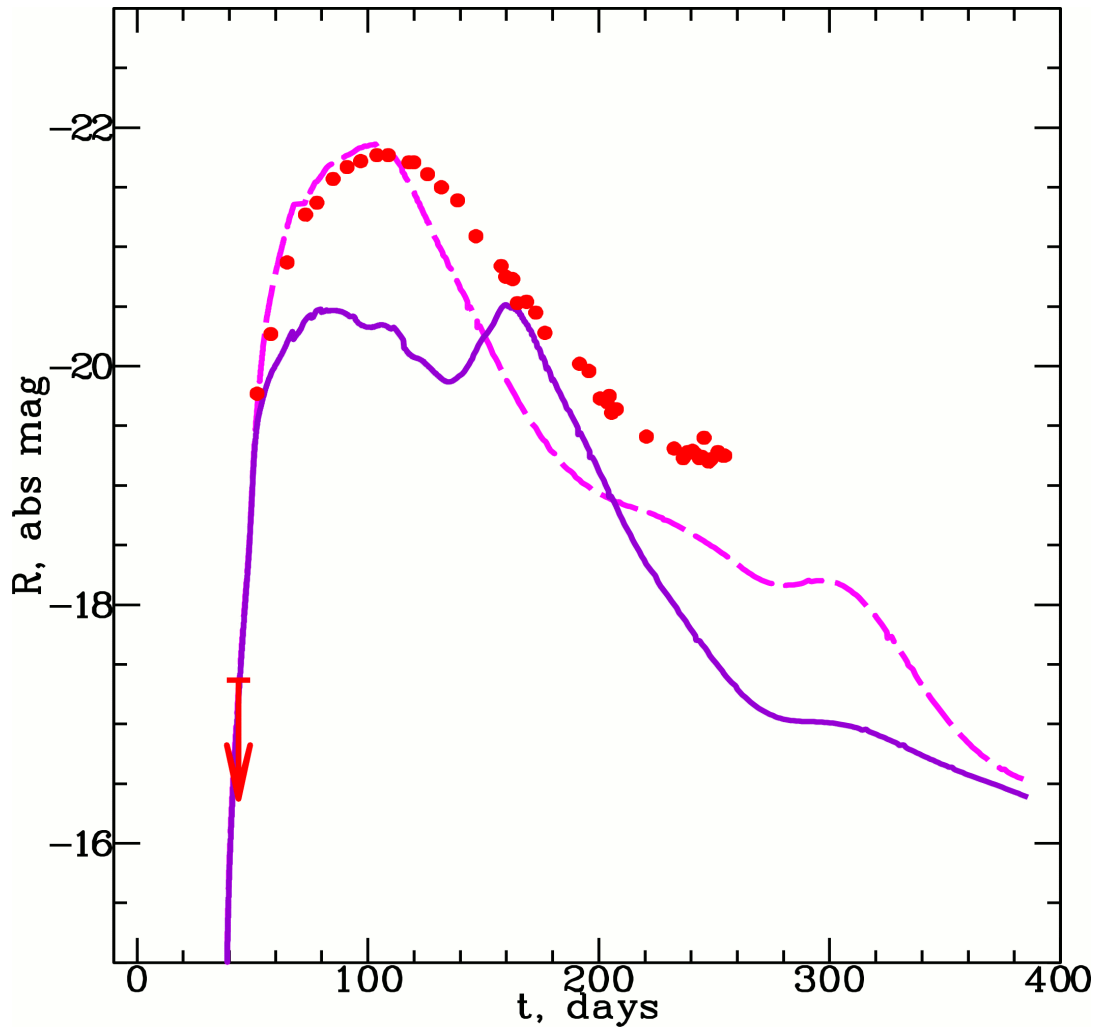

**Figure 3 Absolute R-band magnitudes resulting from the strong second explosion of the 110-solar-mass model.** The time axis has been adjusted so as to give the best agreement with observations of SN 2006gy[1], plotted as the red data points, and the model results have been smoothed using a numerical averaging over 30-day intervals. Multi-dimensional calculations of similar models[31] suggest that instabilities in the thin dense shell, where the radiation originates, will result in the formation of a mixed layer with relative thickness $\Delta R/R \approx 0.1$–$0.15$. The predictions of our one-dimensional model (where the radius is proportional to the time) should thus be blurred by $\Delta t \approx 30$ days for a total light-curve width of about 200 days. An R-band extinction of 1.68 magnitudes is assumed for the supernova[1]. Two curves are shown, one for the nominal model discussed in the text, and a second where the velocity of all the ejecta—pulses 1 and 2—has been multiplied by two (hence an artificial increase in the explosion energy from $7.2 \times 10^{50}$ erg to $2.9 \times 10^{51}$ erg). The large variations apparent in the fainter model are absent in the brighter one because the photosphere at peak light in the more energetic model has not receded to near the shock. The actual explosion energy and mass ejected are sensitive to the initial mass of the star, its uncertain mass loss (that is, the mass of the remaining hydrogen envelope when the star dies), and details of the cooling between pulses. Other light curves, without smoothing and with variable explosion energy and density, are given in Supplementary Figs 7–9.

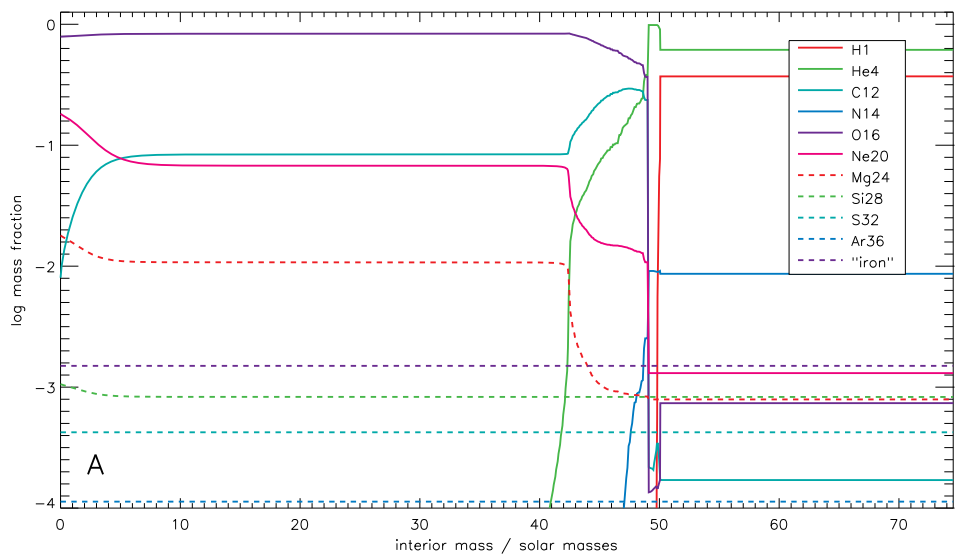

**Supplementary Figure S1.** Initial composition of the 110 solar mass model as it encounters the electron-positron pair instability for the first time. Carbon has already been burned away in the inner few solar masses of the star and the central temperature and density are $1.2 \times 10^9$ K and $9.1 \times 10^4$ g cm$^{-3}$ respectively. Most of the "helium core" is in fact composed of oxygen. The net binding energy of the star at this point (essentially that of the helium-oxygen core) is $4.03 \times 10^{51}$ erg. The total mass of the star is 74.56 solar masses and the helium core is 49.86 solar masses. Tight hydrostatic equilibrium prevails throughout the core, though it is starting to contract rapidly.

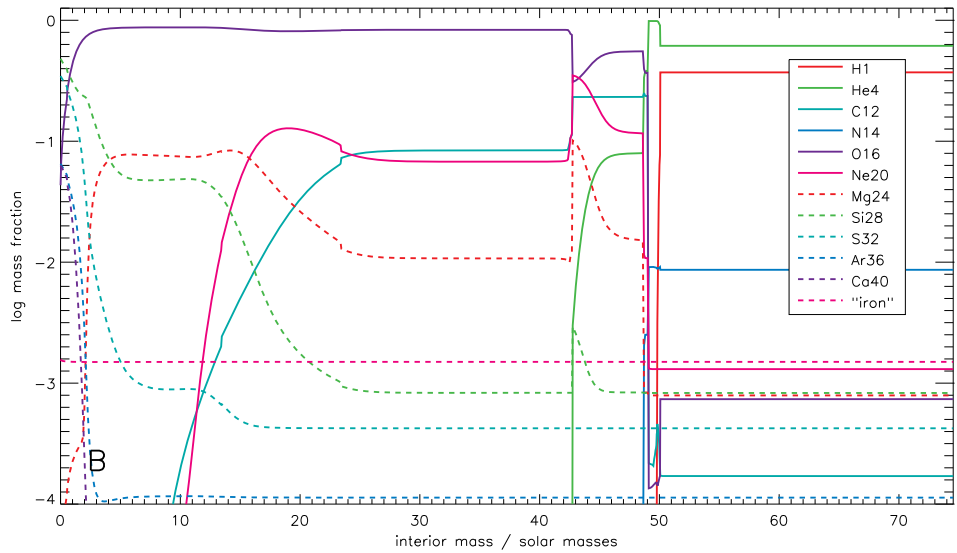

**Supplementary Figure S2.** Composition of the same star as in Fig. S1 at the end of the first pulse. 1.49 solar masses of $^{16}$O and 1.55 solar masses of $^{12}$C have burned and the binding energy is now reduced to $2.66 \times 10^{51}$ erg. The central temperature is $9.25 \times 10^9$ K and the density is $3.34 \times 10^4$ g cm$^{-3}$. The neutrino luminosity of the whole core is $7.6 \times 10^{41}$ erg s$^{-1}$ and this greatly dominates the radiation transport. The mass of the bound core is 50.8 solar masses.

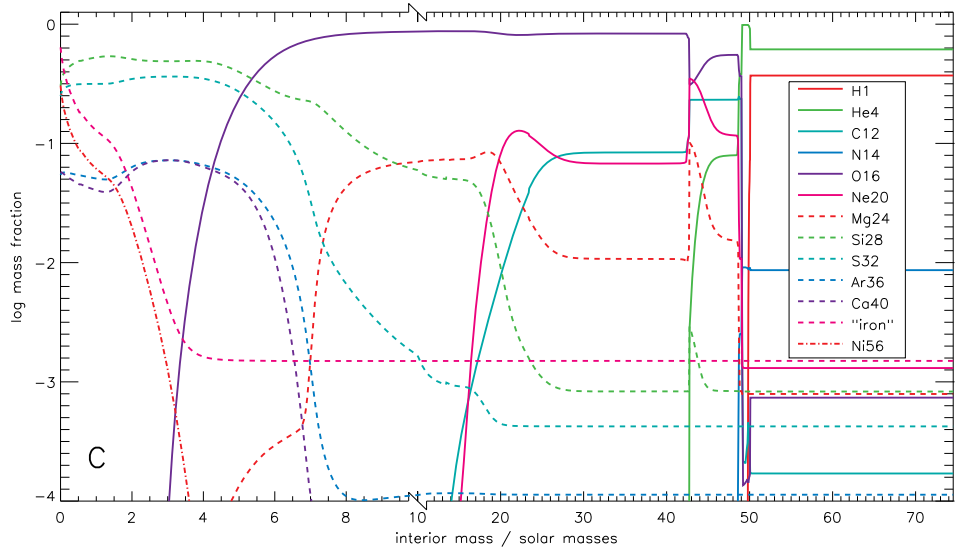

**Supplementary Figure S3.** Composition at the end of the second pulse. Note the scale break in mass at 10 solar masses in this figure and Fig. S4. Another 3.8 solar masses of $^{16}$O and 1.5 solar masses of $^{12}$C have now burned. The stars binding energy is $1.44 \times 10^{51}$ erg and the central temperature and density are $9.60 \times 10^{8}$ K and $2.34 \times 10^{5}$ g cm$^{-3}$. Note the appreciable decrease in entropy (increase in central density at a given temperature) because of neutrino emission during the first interpulse period. The mass of the bound core is now 45.36 solar masses. Because of the reduced entropy and mass, the core does not encounter the pair instability a third time.

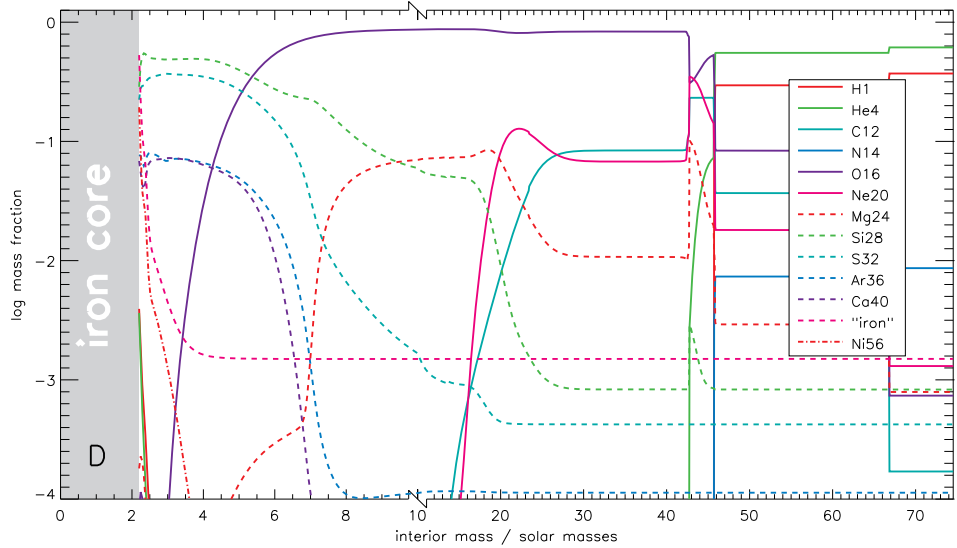

**Supplementary Figure S4.** Composition when the 110 solar mass star finally dies. The iron core mass is 2.18 solar masses and its outer edge is collapsing at 1000 km s$^{-1}$. A hot proto-neutron star will now form, which depending on the maximum physically allowed neutron star mass and accretion over the next few seconds, may become a black hole. A separate calculation of a rotating 95 solar mass star with similar final helium core mass gave an angular momentum for the iron core of $4.3 \times 10^{48}$ erg s, implying a neutron star rotation rate of 2 ms. This is enough angular momentum that rotation is likely to play a role in the final death of the star, perhaps producing a millisecond magnetar or collapsar.

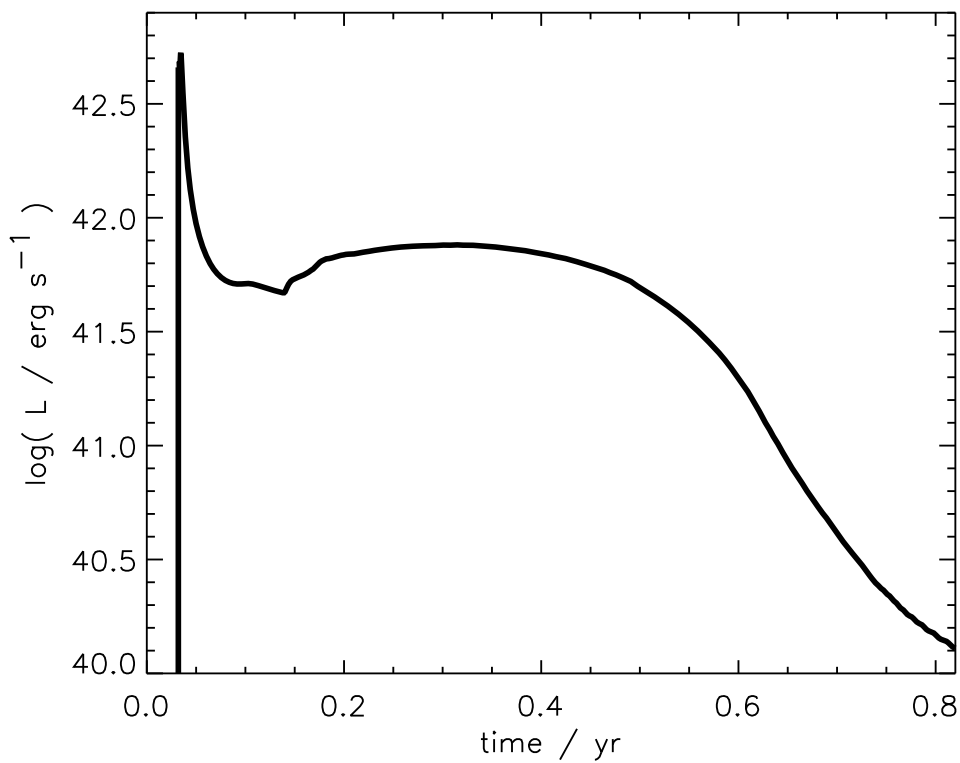

**Supplementary Figure S5.** Light curve resulting from the first mass ejection. The initial spike is from shock wave break out and is not accurately calculated due to coarse surface zoning and the use of a single temperature to describe the radiation and the matter. The roughly 200 day plateau occurs as the hydrogen and helium in the ejected envelope recombine releasing the energy deposited there by the shock. No radioactivity is ejected in either of the pulses, and the light curve has no tail. Zero time here corresponds to shock break out at the surface.

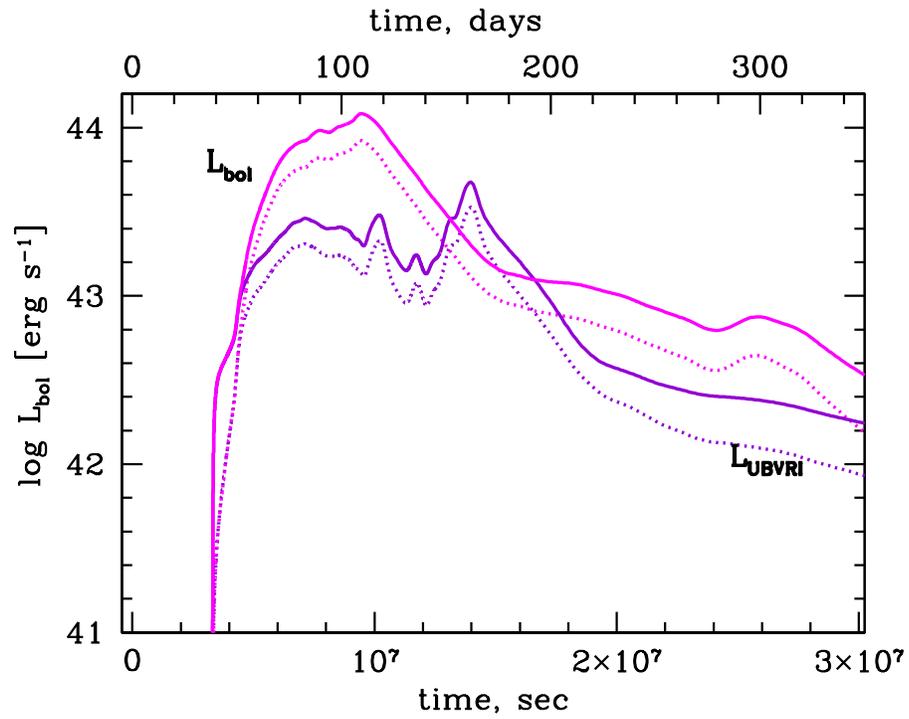

**Supplementary Figure S6.** Bolometric light curves and UVOIR light curves of the standard 110 solar mass model and of the model in which the velocities of both mass ejections were doubled. Zero time here and in subsequent plots is the time at which the second mass eruption began plus 40 days, i.e., the time when the second pair instability led to explosion plus an offset to facilitate comparison with SN 2006gy.

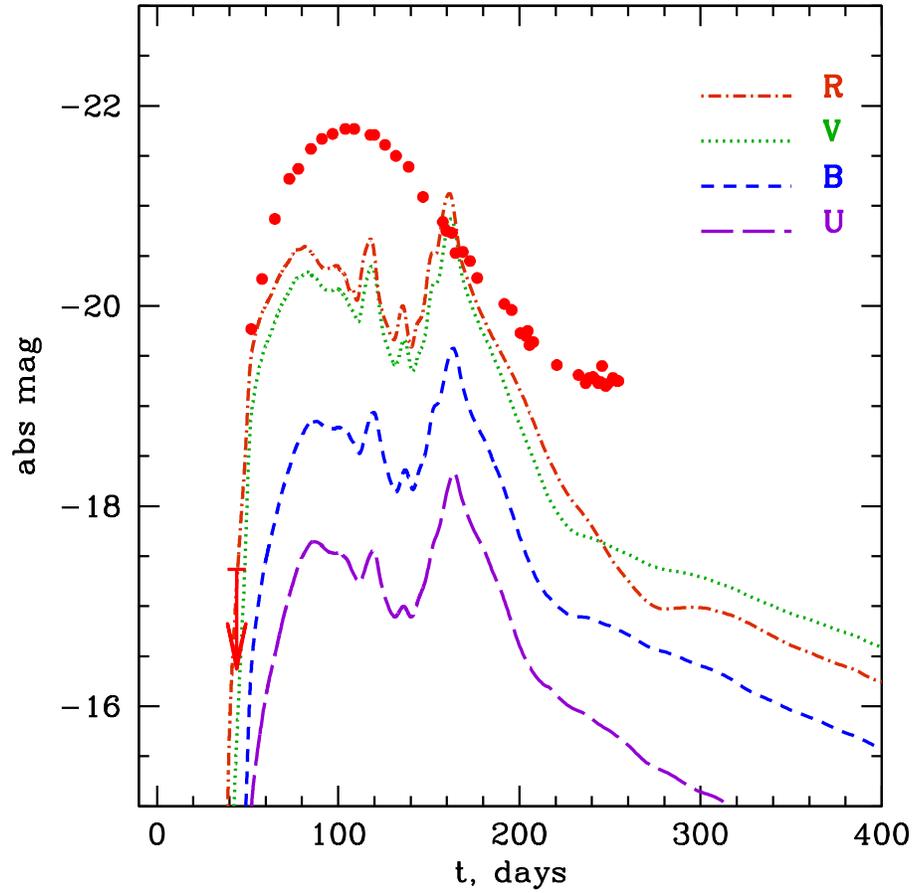

**Supplementary Figure S7.** Color magnitudes of the 110 solar mass model. The supernova is brightest in the red and visual bands. Unlike Fig. 3 in the main text, the light curve has not been smoothed, but shows the rapidly varying temporal structure resulting from an artificial 1D simulation of thin shells. In two or three dimensions, the thin shell is expected to break up into multiple structures at different radii. An oblateness in either mass ejection would also smooth out much of this time structure. In this and subsequent plots, an extinction in the R-band of $A_R = 1.68$ magnitudes has been assumed in plotting the data for SN 2006gy. Zero time is the time of the second pulse plus 40 days.

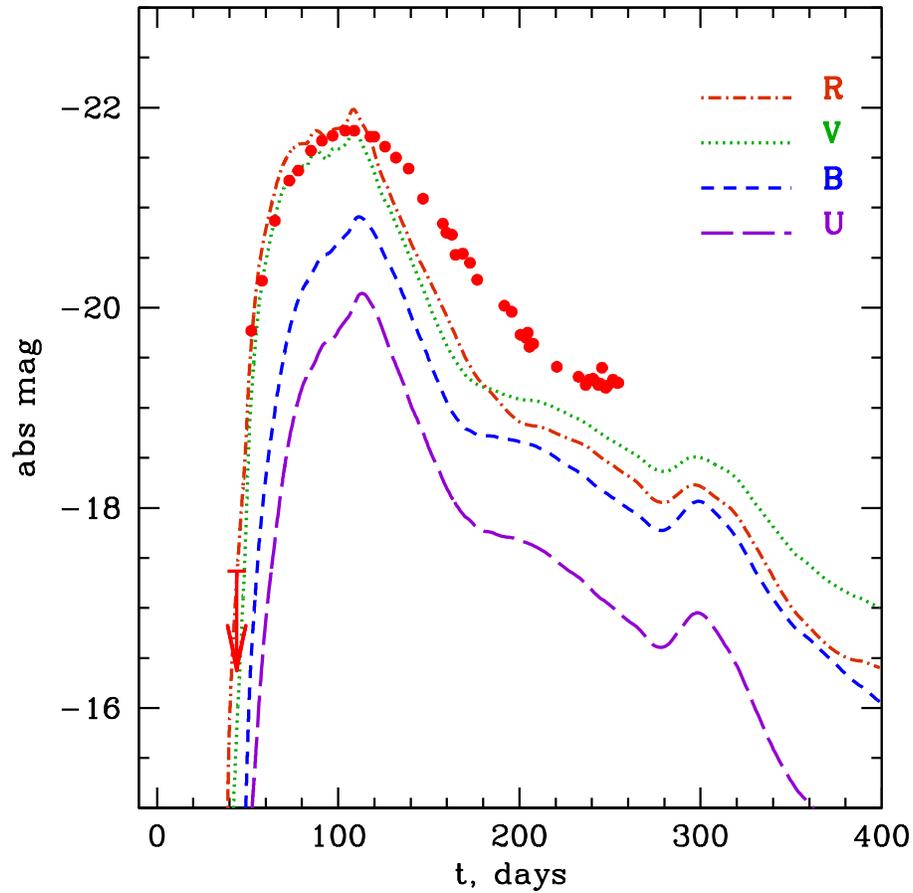

**Supplementary Figure S8.** Color magnitudes of a version of the 110 solar mass model in which the velocity of all the ejecta was multiplied by two. This requires a quadrupling of the total explosion energy to $2.9 \times 10^{51}$ erg, which is rather large compared with the pulses for a variety of models in Supplemental Table S1. A smaller increase in velocity is needed if the density (i.e., mass ejected) is also increased (Fig. S9). Actually, just doubling the velocity of the second ejection gives virtually the same result, so the real requirement here is an energy for the second pulse of $\sim 2 \times 10^{51}$ erg (or a reduction in the observed brightness of SN 2006gy).

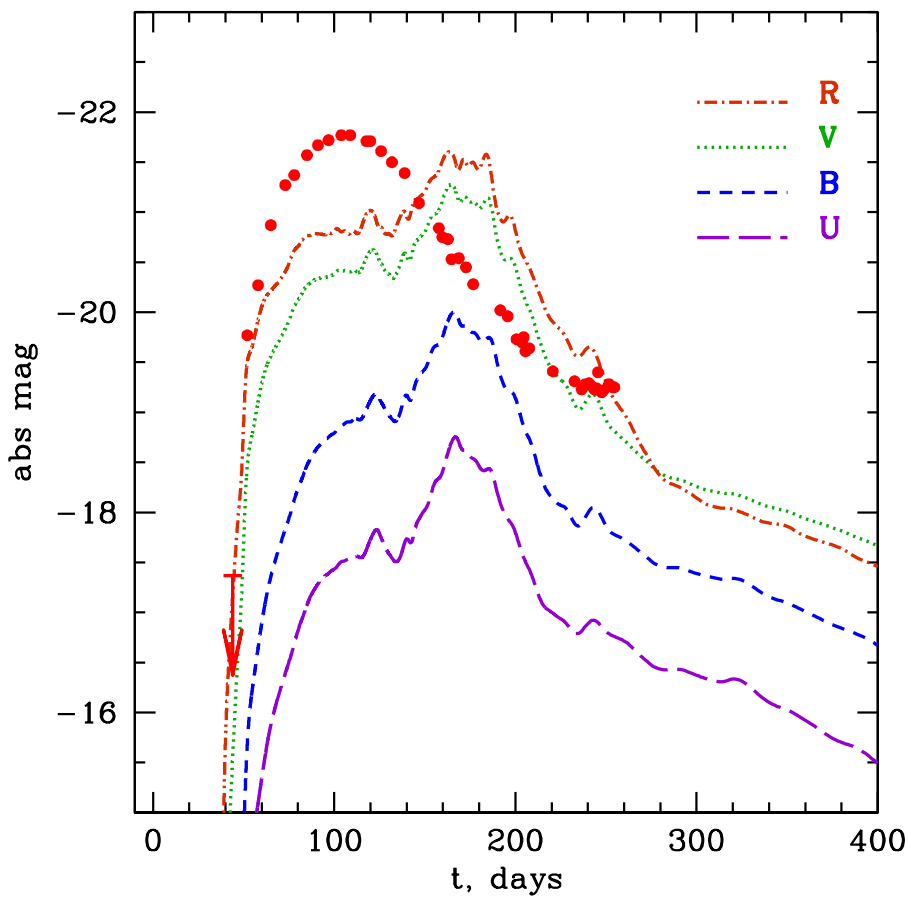

**Supplementary Figure S9.** Color magnitudes of the 110 solar mass model in which the density of the both pulses was multiplied by two. This doubles both the amount of mass ejected in each pulse and the total kinetic energy. Some combination of increased energy (Fig. S8) and density (this figure) could probably be found that would fit the SN 2006gy observations.

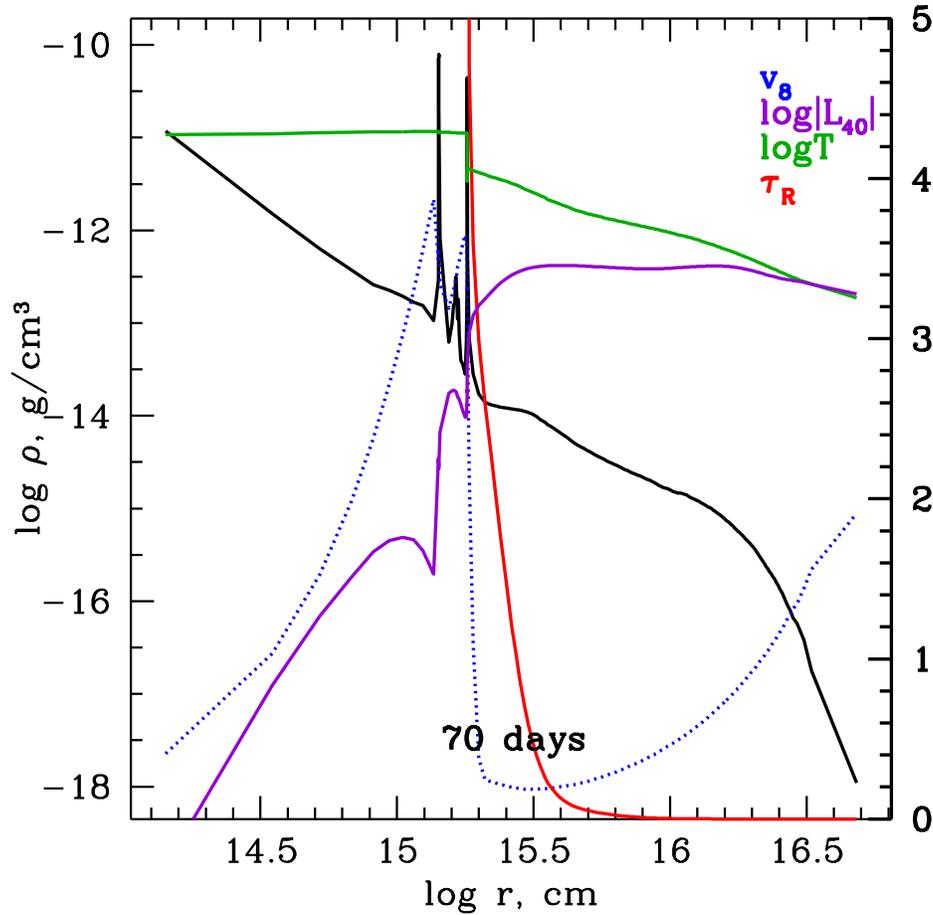

**Supplementary Figure S10.** Photospheric structure at 70 days. Plotted are the variation in log density (g cm-3), velocity (1000 km s-1), log temperature (K), log luminosity ($10^{40}$ erg s-1, and optical depth ($\tau_R$) in the emitting region for the 110 solar mass model. Because of the higher density and smaller radius at this early time, the temperature in the vicinity of the shock is higher, keeping the matter ionized for some distance ahead. Consequently, the photosphere is well outside the shock and the emission is black body. No x-rays are being created because the temperature is too low and none would escape if they were because of the large column depth.

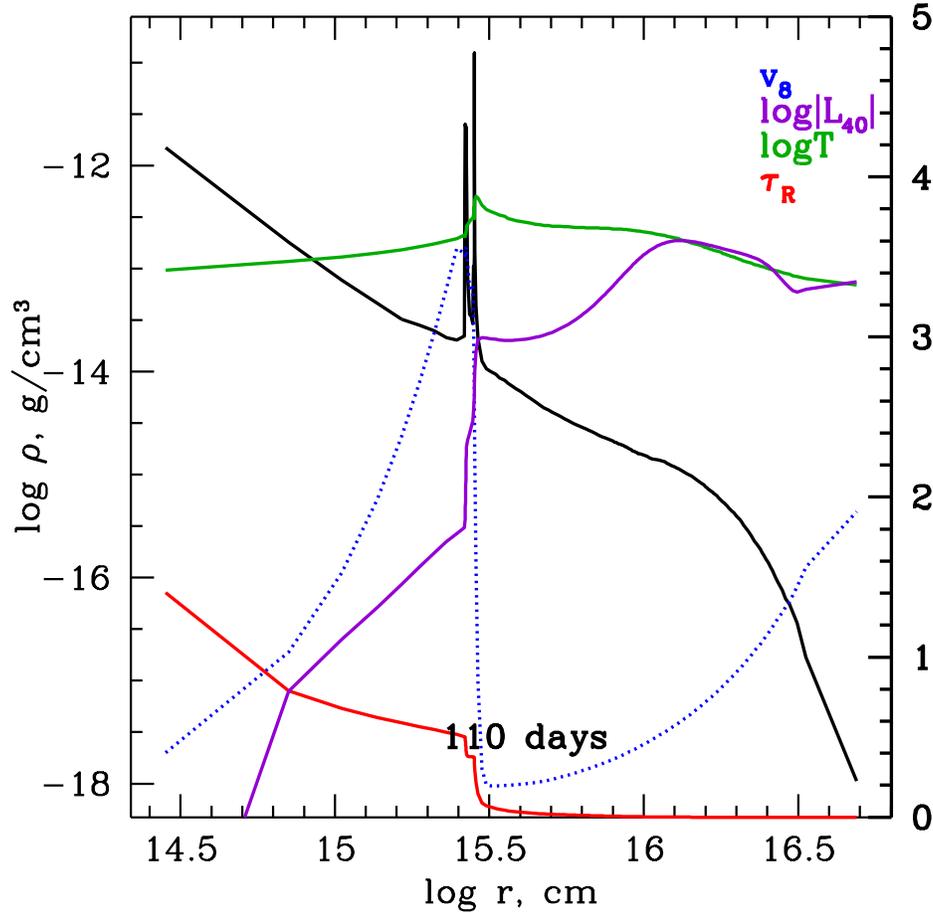

**Supplementary Figure S11.** Photospheric structure at 110 days. Similar quantities are plotted to Fig. S8, but for a later time near peak light. The photosphere has now receded almost to the shock, but appreciable opacity is maintained by the Doppler broadened atomic lines. $\tau_R$ is Rosseland opacity which, in the outer regions where the light originates, is controlled by metal lines, but in hot regions still has an appreciable contribution of Thomson scattering. The optical depth to x-rays remains high but perhaps not so high that a little x-ray emission couldn't leak out. More problematic is the lack of any physical mechanism in the Stella code for generating non-thermal electrons. However, the physics in the code does predict the predominance of optical-IR emission for a long time after peak.

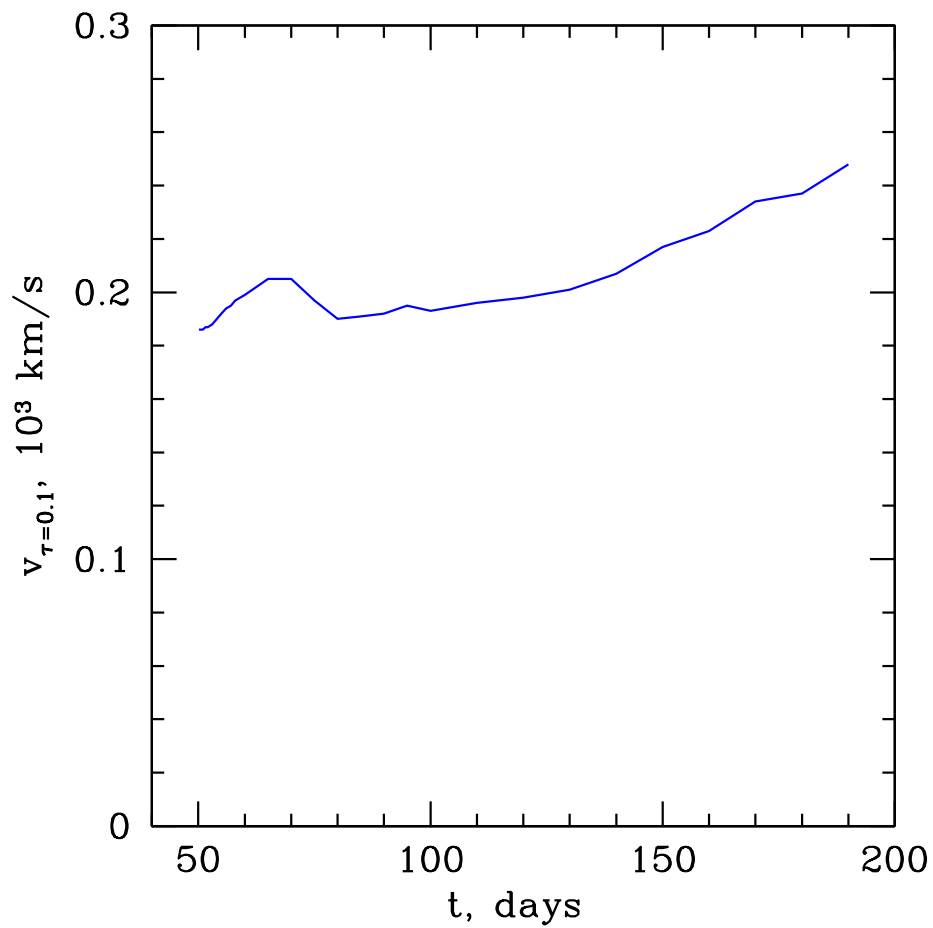

**Supplementary Figure S12.** Velocity at optical depth $\tau = 0.1$, as might be typical for the narrow lines seen in the spectrum of a Type IIn supernova. The characteristic velocity remains near 200 km s$^{-1}$ throughout the peak of the light curve but rises gradually with time.

**Supplementary Table 1.** The table on the following page summarizes the outburst history for non-rotating helium stars (initially 98.5% helium and 1.5% nitrogen) of various masses evolved to the point of final central collapse. The main sequence masses corresponding to these cores are approximately 2.2 times the helium core mass, i.e., 105 to 130 solar masses. Within this range, stars with larger mass encounter an instability that is increasingly violent and a smaller number of pulses occur before the star dies. In each case, the first pulse is the weakest but, except in the lightest case considered, more than adequate to eject any residual hydrogen envelope on the first try. The kinetic energy of the remaining pulses is larger, though never much over $10^{51}$ erg. This energy is no longer shared with any envelope and thus has a higher velocity. Each pulse will thus produce a supernova-like display. The supernovae may be exceptionally brilliant, as in the case considered here, or quite faint for stars on the lower end of the unstable mass range. The table gives, for each pulse, the kinetic energy, (KE), and mass ejected ($\Delta$M). Following each pulse and a brief period of oscillation, the core has central temperature, $T_c$, and density, $\rho_c$, and radiates neutrinos for the interval given before encountering the next instability. Each interval is given in the form of a number and the power of ten in parentheses by which that number is to be multiplied. At the end of the last pulse listed, the central part of the core evolves to an iron core that collapses to a neutron star or black hole. Helium cores of increasing mass encounter an instability that is increasingly violent on the first encounter. For helium cores above 65 solar masses the entire star is disrupted in a single flash. For stars with hydrogenic envelopes the evolution is altered since the helium core can grown by hydrogen shell burning even as it burns helium in the center. The 110 solar mass model discussed in the text is similar to the 51 solar mass helium core model above, but only experienced two strong outbursts before dying.

| He Mass $(M_\odot)$ | Pulse | $KE_1$ $(10^{50}$ erg$)$ | $\Delta M$ $(M_\odot)$ | $T_c$ $(10^9$ K$)$ | $\rho_c$ $(10^5$ g cm$^{-3})$ | interval (sec) |
|---|---|---|---|---|---|---|
| 48 | 1 | 0.048 | 0.11 | 1.48 | 1.68 | 7.34(5) |
|    | 2 | 0.92  | 0.57 | 1.57 | 2.02 | 4.31(5) |
|    | 3 | 2.20  | 1.19 | 1.31 | 1.34 | 2.77(6) |
|    | 4 | 3.09  | 1.64 | 1.38 | 3.00 | 2.02(6) |
|    | 5 | 4.41  | 1.84 | 1.32 | 3.40 | 8.33(6) |
|    | 6 | 3.02  | 2.42 | 1.86 | 28.6 | 7.43(5) |
| 51 | 1 | 0.26  | 0.44 | 1.17 | 0.67 | 1.02(7) |
|    | 2 | 2.70  | 1.55 | 1.30 | 1.80 | 2.72(6) |
|    | 3 | 4.49  | 1.99 | 1.06 | 1.66 | 2.74(7) |
|    | 4 | 7.56  | 3.68 | 1.22 | 3.77 | 2.53(7) |
| 52 | 1 | 0.85  | 1.13 | 1.01 | 0.40 | 6.32(7) |
|    | 2 | 1.46  | 0.94 | 1.57 | 5.02 | 4.58(5) |
|    | 3 | 4.27  | 1.90 | 1.16 | 2.74 | 8.10(6) |
|    | 4 | 7.29  | 3.12 | 1.09 | 2.68 | 9.56(7) |
| 54 | 1 | 3.11  | 3.23 | 0.71 | 0.14 | 6.13(9) |
|    | 2 | 2.51  | 2.09 | 1.57 | 14.6 | 8.85(5) |
|    | 3 | 5.33  | 2.68 | 1.01 | 3.33 | 3.73(8) |
| 56 | 1 | 2.44  | 2.71 | 0.74 | 0.15 | 3.47(9) |
|    | 2 | 1.45  | 1.34 | 1.57 | 8.7  | 4.32(5) |
|    | 3 | 6.12  | 3.33 | 1.03 | 3.02 | 1.44(8) |
| 58 | 1 | 13.3  | 9.39 | 0.24 | 0.0072 | 1.24(11) |
|    | 2 | 4.00  | 2.39 | 1.46 | 6.08 | 2.10(6) |
|    | 3 | 7.78  | 3.06 | 1.07 | 3.31 | 1.61(8) |
| 60 | 1 | 20.6  | 17.6 | 0.087 | 0.0004 | 1.86(11) |
|    | 2 | 1.17  | 0.78 | 1.77 | 10.2 | 2.90(5) |